\documentclass[%
superscriptaddress,
 reprint,
 amsmath,amssymb,
 aps,
]{revtex4-2}

\usepackage{graphicx}
\usepackage{dcolumn}
\usepackage{bm}
\usepackage{xurl}

\begin{document}


\title{{Observation of narrow-band $\gamma$ radiation
from a boron-doped diamond superlattice with an 855 MeV electron beam}}


\author{Hartmut Backe}
\email[Corresponding author: ]{backe@uni-mainz.de}
\affiliation{{Institute for Nuclear Physics of Johannes Gutenberg-University}, {Johann-Joachim-Becher-Weg 45},{Mainz},{D-55128},{Germany}}

\author{José Baruchel}
\affiliation{{European Synchrotron Radiation Facility},{71 Avenue des Martyrs},{Grenoble},{F-38000},{France}}

\author{Simon Bénichou}
\affiliation{{European Synchrotron Radiation Facility},{71 Avenue des Martyrs},{Grenoble},{F-38000},{France}}

\author{Rébecca Dowek}
\affiliation{{European Synchrotron Radiation Facility},{71 Avenue des Martyrs},{Grenoble},{F-38000},{France}}

\author{David Eon}
\email[Contact author:]{ david.eon@neel.cnrs.fr}
\affiliation{{Université Grenoble Alpes},{Institut Néel, CNRS UPR2940},
{Grenoble cedex 9},{F-38042},{France}}

\author{Pierre Everaere}
\affiliation{{European Synchrotron Radiation Facility},{71 Avenue des Martyrs},{Grenoble},{F-38000},{France}}

\author{Lutz Kirste}
\affiliation{{Fraunhofer Institute for Applied Solid State Physics (IAF)},{Tullastrasse 72},{Freiburg},{D-79108},{Germany}}

\author{Pascal Klag}
\affiliation{{Institute for Nuclear Physics of Johannes Gutenberg-University}, {Johann-Joachim-Becher-Weg 45},{Mainz},{D-55128},{Germany}}

\author{Werner Lauth}
\email[Contact author: ]{lauthw@uni-mainz.de}
\affiliation{{Institute for Nuclear Physics of Johannes Gutenberg-University}, {Johann-Joachim-Becher-Weg 45},{Mainz},{D-55128},{Germany}}

\author{Patrik Straňák}
\affiliation{{Fraunhofer Institute for Applied Solid State Physics (IAF)},{Tullastrasse 72},{Freiburg},{D-79108},{Germany}}

\author{Thu Nhi {Tran Caliste}}
\email[Contact author: ]{thu-nhi.tran-thi@esrf.fr}
\affiliation{{European Synchrotron Radiation Facility},{71 Avenue des Martyrs},{Grenoble},{F-38000},{France}}


\date{\today}

\begin{abstract}
We report the first observation of narrow band 1.3 MeV $\gamma$ radiation from a crystalline diamond micro-undulator. A diamond superlattice was grown with a periodical varying boron doping profile. Four sinusoidally deformed (110) periods resulted with a period length of 5.0 $\mu$m and an amplitude of 0.098 nm. A channeling experiment was performed with the 855 MeV electron beam of the Mainz Microtron MAMI accelerator facility. A clear peak was detected with a large sodium iodide scintillation detector close to the expected photon energy of 1.28 MeV. Key characteristics of the peak, including photon energy, width and intensity, were reproduced fairly well by Monte-Carlo simulation calculations. Based on the latter, optimized boron doping profiles were designed for a hypothetical 3 GeV electron beam, enabling preparation of a highly directional $\gamma$-ray beam with a photon energy of 14.5 MeV. The predicted spectral bandwidth is 13\%, however, with a high energy continuum tail. The on-target photon flux at a beam current of 100 $\mu$A would be about $10^{12}$/s.
\end{abstract}


\maketitle

\section{Introduction}\label{intro}
\noindent Intense narrow-band multi-MeV photon beams hold significant potential for applications in basic and nuclear research, medicine and industrial technology \cite{KleP25,ZilB22,IAEA-TECDOC,HowA22,BudB22}. Having in mind in particular photonuclear reactions for application purposes, the production mechanisms of such beams should be targeted, from an economical point of view, on electron accelerators with moderate energies in the GeV range or below. Bremsstrahlung radiation created by an electron beam incident on a high-Z target has been widely used for many decades. However, such sources feature broad band radiation characteristics. Nearly mono-energetic $\gamma$-ray beams of high intensity can be produced by facilities based on laser Compton-backscattering which are already operational \cite{WelW09} or currently under construction \cite[Table 10]{HowA22}.

\begin{figure}[t]
\centering
    \includegraphics[angle=0,scale=0.70,clip]{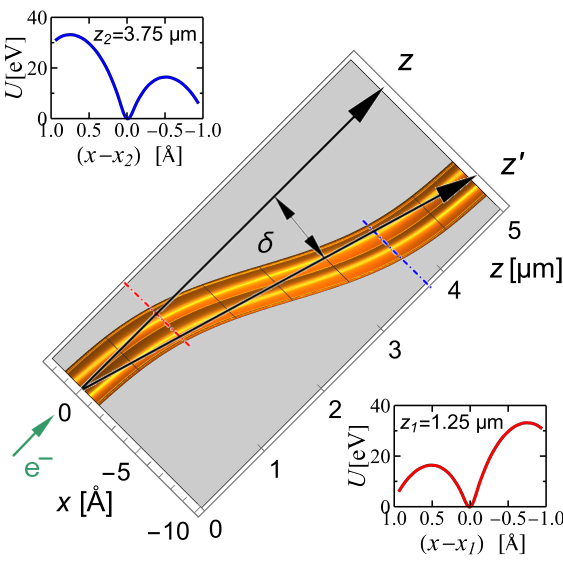}
\caption[]{Transverse potential $U(x,z)$ for a single (110) channel of a boron doped diamond superlattice. Shown is one period  with length of $\lambda$ = 5.0 $\mu$m.
Coordinates $(x,z)$ of a full period (0,0) and (-7.50 \AA, 5 $\mu$m) define the $z'$ undulator direction around the potential wiggles. It makes an angle $\delta$ = -0.150 mrad with $z$ axis, the nominal electron beam direction. Notice, the longitudinal $z$ coordinate is squeezed by four orders of magnitudes.
For details see appendix \ref{simulation} and also \cite{KorS26}. Barrier heights are modified as functions of $z$ by centrifugal forces. Insets show examples, $x_1$ = -0.898 {\AA} and $x_2$ = -6.604 {\AA}. Figure based on design parameters of section \ref{diamondUndulator}.
} \label{potentialBent}
\end{figure}

In this work an alternative route towards narrow band photon sources was elaborated which is based on the channeling process of ultra-relativistic electrons in periodical bent crystals. The method, theoretically known since long \cite{KapK80,BarD80}, has been  comprehensively treated in \cite{KorS04,BelM06,BarT13,SusK13} or in monographs \cite{Bar12,KorS14,KorS22} in which all relevant references can be found. It resembles magnetic undulators, however, with period lengths in the multi-$\mu$m range rather than on a centimeter scale.
The principle will be explained by means of Fig. \ref{potentialBent}. We employ the continuum potential picture introduced by Lindhard \cite{Lin65}, applied also in our case-study \cite{BacL25} on which this publication is partly based.

An electron entering the crystal at the coordinates $x = z$ = 0 nearly parallel to the $z$ axis may be captured in the potential pocket, which at that point is symmetrical with a barrier height of 24 eV.
The electron oscillates while following the bent sinusoidal shape of the (110) plane. By means of the accompanied acceleration channeling radiation and undulator-like radiation will be emitted. The radiation is strongly forward directed, for undulator-like radiation the typical opening angle is in the order of 0.5 mrad for 1 GeV electrons. The photon energies fall into the $\gamma$-ray domain, for a period length of a few $\mu$m typically in the MeV range. It scales with the electron beam energy squared and inversely with the undulator period length $\lambda_U$. The capture probability and in turn the intensity of the $\gamma$ radiation requires high quality electron beams with a low angular spread. The reason is that the transverse energy  $pv/2 \cdot \vartheta_x^2$ is the initial kinetic energy of the particle in the potential pocket with $pv$ in essence the electron beam energy and $\vartheta_x$ the angle of the electron with respect to the $z$ axis.

The experimental challenge is to fabricate crystalline undulators with such small periods. A graded composition strained layer in a superlattice was proposed \cite{IkeL84,MikU00}, and Si$_{1-x}$Ge$_x$ undulator crystals were grown at the Aarhus University. Experiments were performed with an undulator with period length of 9.9 $\mu$m \cite{BacK12,BacK12A} with 270, 375, and 855 MeV electrons at MAMI. Structures at predicted undulator energies were observed and interpreted as originating from synchrotron-like radiation. In \cite{BarT13} Monte-Carlo calculations were performed for the case of 855 MeV which may corroborate this interpretation. Experiments with an undulator with a period length of 0.43 - 0.44 $\mu$m  \cite{WisA14,WisU17} were not convincing enough to be used for applications discussed in \cite{UggW15}, and this route will not be pursued in this paper.
An alternative are diamond undulators which can be produced by Chemical Vapour Deposition (CVD) with periodically varying content of boron. The effect is based on the functional dependence of the lattice constant on the boron content, see e.g. \cite[Fig. 1]{WojA08}.
- Previous attempts to observe undulator radiation from boron-doped diamond structures were unsuccessful \cite{BosC16,BacL25}.

In this paper we report on the first successful observation of narrow band undulator radiation from a boron doped diamond superlattice. The parameters of the micro-undulator, which are quite different in comparison to that reported in \cite{BacK12,BacK12A} for the Si$_{1-x}$Ge$_x$ undulators, are collected in the caption of Fig. \ref{potentialBent}. At the design of the undulator we took into account that the low $Z$ = 6  material diamond has the advantage, beside its radiation hardness, that scattering by atomic nuclei and electrons comprising the crystal is much smaller in comparison with silicon, $Z$ = 14. Scattering may increase the transverse energy and, in turn, the de-channeling probability. This is the case, in particular, in regions in which the radius of curvature is small, i.e., centrifugal forces are large and one of the barrier heights is low as shown in Fig. \ref{potentialBent} at $z_1$ = 1.25 $\mu$m and $z_2$ = 3.75 $\mu$m. Therefore, the radius of curvature was kept large resulting in a small oscillation amplitude or, in other words, in a small undulator $K$ parameter resulting in radiation with moderate intensity.

Applications of such $\gamma$ radiation sources may be placed in the field like nuclear structure research with photonuclear reactions or many others as compiled in \cite{HowA22}.

The paper is organized as follows. In section \ref{experimental}  the superlattice is described, including diagnosis of it, and the experimental setup at MAMI. Section \ref{measurements} is devoted to the peak search. The result is discussed in section \ref{discussion} by means of a comparison with Monte-Carlo simulation calculations. In the following section \ref{prospects} beam features are sketched for possible future applications.
The paper closes in section \ref{Conclusions} with conclusions. Some details of the superlattice preparation as well as of the Monte-Carlo simulation calculations are moved into the appendices \ref{appendices} A and B, respectively.

\section{Experimental} \label{experimental}
\subsection{The diamond undulator} \label{diamondUndulator}
\noindent Parameters of the diamond superlattice were optimized by means of Monte-Carlo simulation calculations for an experiment with an electron beam energy of 855 MeV. The period length must be chosen such that the undulator peak energy is well separated from the channeling radiation distribution. Finally we found the following design parameters: period length in [100] direction 3.54~$\mu$m, number of periods $N_U$ = 4, and amplitude $A_U$ = 0.098 nm.
Details are described in appendix \ref{superlattice}.
\begin{figure}[h]
\centering
\includegraphics[angle=0,scale=0.35,clip]{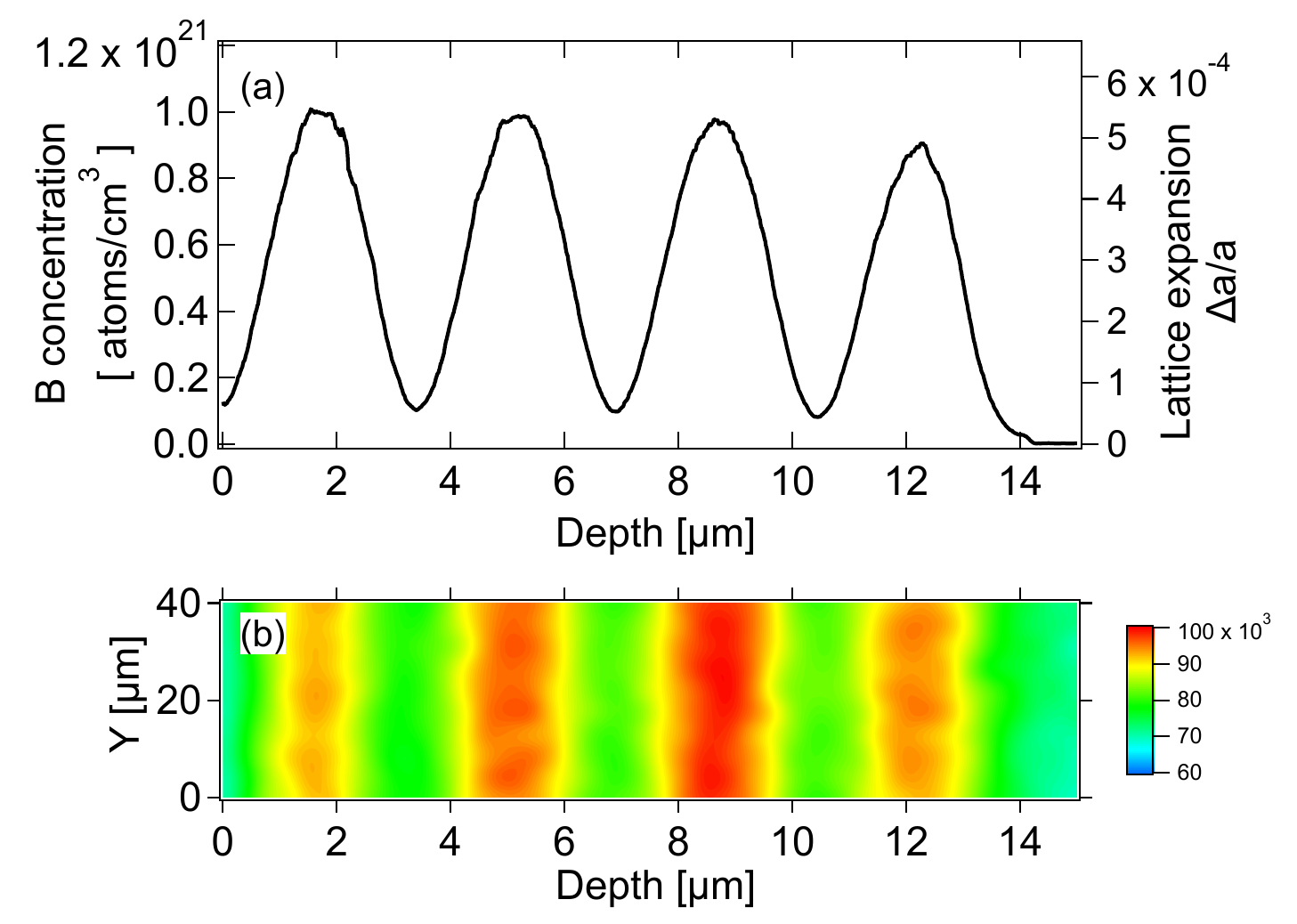}
\caption[]{(a) Measured boron doping profile. A 250 x 250 µm² crater allowed for the analysis of the material as a function of depth with ToF-SIMS. On the left scale the boron concentration is shown, on the right one the corresponding relative lattice expansion $\Delta$a/a, according to the "atomic volume interpolation" \cite{WojA08}, also called "modified Vegard’s law".
(b) Depth profile obtained with the Rocking Curve Imaging (RCI) technique \cite{CalK23, BarE26}. The signal is the integrated intensity of Bragg-diffraction peaks  at (111) planes with 17 keV photons.}
\label{SIMSBraggMeasurement}
\end{figure}

The boron doping profile was examined with the Time-of-Flight Secondary Ion Mass Spectrometry (ToF-SIMS) method. Fig. \ref{SIMSBraggMeasurement} (a) shows the depth profile. The ToF-SIMS data indicate that the growth process reproduce the four sinusoidal periods, irrespective of the analyzed region. As designed, the layers are consistent with a period length of ~3.54 $\mu$m along the [100] growth direction.
A slight variation of the doping maxima and minima from one period to the other can be observed for the layers apart from the substrate.

Fig. \ref{SIMSBraggMeasurement} (b) depicts a section of an integrated intensity map of the Bragg peak, taken with the Rocking Curve Imaging (RCI) technique  \cite{CalK23, BarE26} at the European Synchrotron Radiation Facility, Grenoble. The structure  corresponds to the doping profile of Fig. \ref{SIMSBraggMeasurement} (a). This result has been confirmed over more than 1 mm along the crystal.
\begin{figure}[b]
\centering
  \includegraphics[angle=0,scale=0.5,clip]{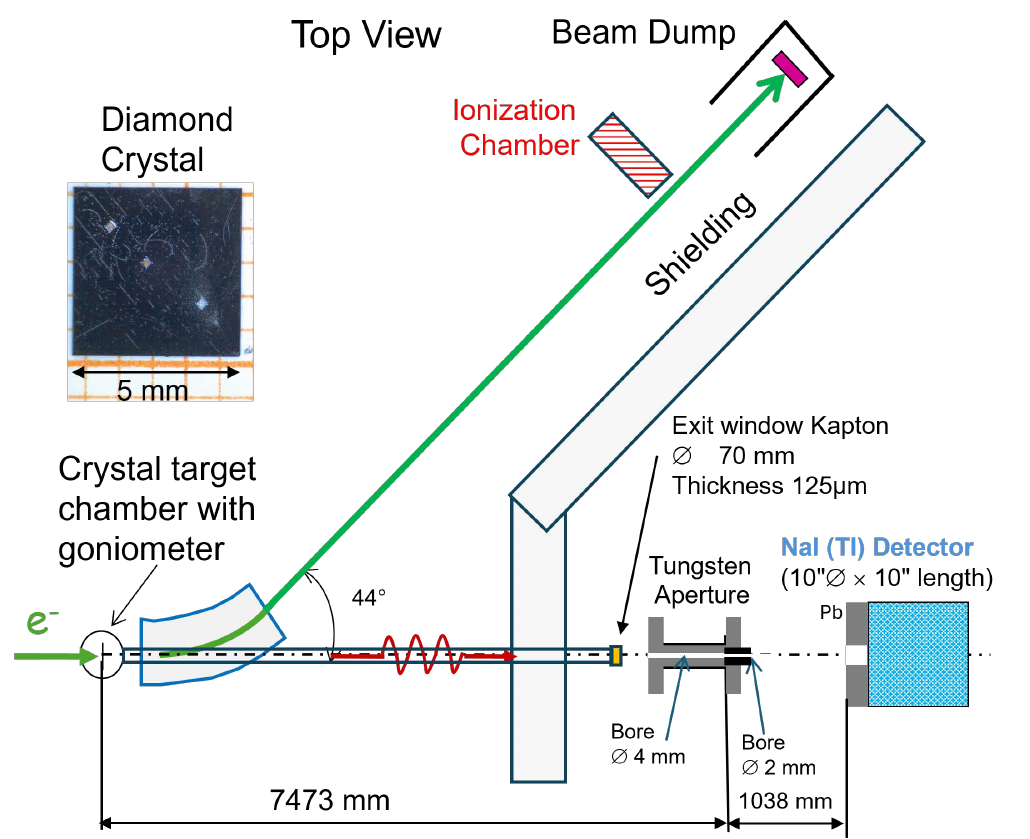}
\caption[]{Experimental setup at MAMI, not to scale. Radiation can be observed at virtual collinear geometry of the electron beam direction and the sodium iodide NaI(Tl) radiation detector. The inset depicts the undulator crystal with the position of three ToF-SIMS craters.}\label{setup}
\end{figure}

\subsection{Experimental setup at MAMI} \label{AcceleratorConfiguration}
\noindent The experimental setup at the Mainz Microtron MAMI is shown at Fig. \ref{setup}. The electron beam is focused onto the diamond crystal in the target chamber with a goniometer, and thereafter deflected by a dipole magnet into the beam dump. The target can be rotated in all 3 spatial directions with an accuracy of 35~$\mu$rad. The target is aligned using the signal of the ionization chamber that registers scattered electrons. The orientation of the electron beam parallel to planes increases the scattering probability and, in turn, the signal of the ionization chamber. This method allows rapid alignment to a crystal plane.

The radiation in the forward direction is detected with a 10 inch sodium iodide crystal, NaI(Tl), positioned at a distance of about 8.5 m from the target. A movable cylindrical tungsten aperture defines the observation direction. It has a length of 261 mm and a bore of 4~mm diameter which was reduced with an additional 28.8~mm long tungsten cylinder inset with an inner diameter of 2~mm. The alignment of the aperture was done by maximizing the detected bremsstrahlung from an aluminum foil.
\begin{figure}[b]
\centering
\hspace*{1.5 mm}
 \includegraphics[angle=0,scale=0.34,clip]{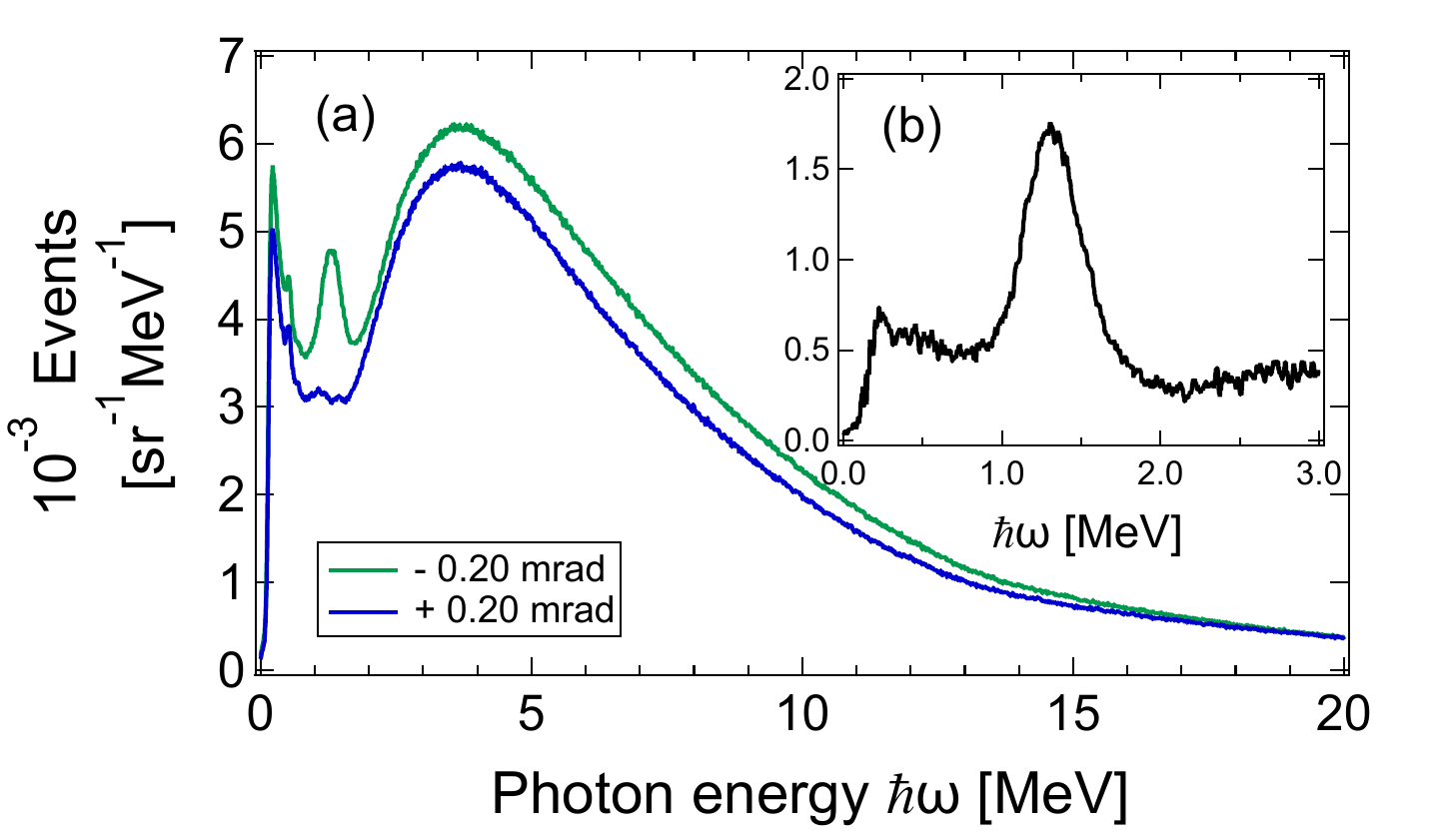}
 \includegraphics[angle=0,scale=0.66,clip]{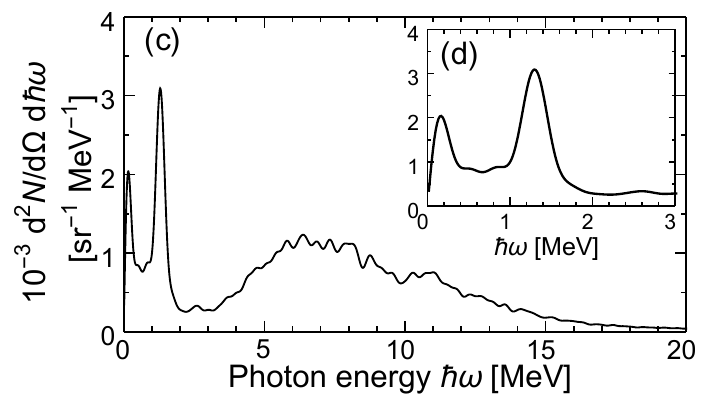}
\caption[]{(a) Experimental spectra of the undulator chip per electron, including the 182 $\mu$m thick backing crystal. Spectra taken with the NaI(Tl) detector at a typical beam current of 3.95~pA. Spectrum in green colour for detuning the aperture from $z$ direction by $\theta_x$ = - 0.2 mrad, and in blue colour by + 0.2 mrad. Both are dominated by channeling radiation of the backing crystal.
(b) Low energy part of difference spectrum. Background spectrum taken at $\theta_x$ = +0.2 mrad aperture position (blue) subtracted from spectrum taken at -0.2 mrad (green).
(c) and (d) Monte-Carlo simulated spectra per electron without a backing crystal, including scattering off atoms and electrons comprising the crystal \cite{BacL25,Bac22A}. Ideal sinusoidal boron doping profile and infinitesimal small aperture size assumed, beam divergence $\sigma_x = \sigma_y$ =0.015 mrad. Mean of 400 single simulations.}
\label{ExpPeakSearch}
\end{figure}

\section{Measurements and results} \label{measurements}
\noindent The peak search was performed at an electron energy of 855 MeV. The angle of the tungsten aperture was varied. There exist two favourable observation angles. One is the direct forward direction $\theta_x = 0$ for which the radiation direction defining aperture is on-axis with the electron beam direction. It is in Fig. \ref{potentialBent} the $z$ axis which makes a small angle $\delta_0$ = 0.0272 mrad with the [110] direction of the host crystal due to the small residual boron doping.
The other one is the undulator direction in which the aperture is detuned by the angle $\theta_x=\delta$ = - 0.150 mrad into the $z'$ direction. For both directions $\theta_y = 0$ was chosen. For details see appendix \ref{superlattice}.

The undulator-photon energy can be estimated with the well known undulator relation
\begin{eqnarray} \label{undulatorEnergy}
\hbar\omega=k\frac{4\pi\gamma^2\hbar c}{\lambda_U \left(1+K^2/2 + \gamma^2 (\delta-\theta_x)^2 + (\gamma \theta_y)^2
\right)}
\end{eqnarray}
with Lorentz factor $\gamma$ = 1674.2 for 855 MeV,
the reduced Planck constant $\hbar$, and $c$ the speed of light. In lowest order $k$ = 1 a peak energy $\hbar\omega$ = 1.28 MeV results
at observation in undulator $z'$ direction $\theta_x$ - $\delta$ = 0, $\theta_y$ = 0. For the undulator amplitude $A_U$ = 0.0977 nm, $\lambda_U$ = 5.0 $\mu$m, the undulator parameter amounts to $K = \gamma\cdot A_U \cdot 2\pi/\lambda_U$ = 0.206.

Fig. \ref{ExpPeakSearch} depicts spectra taken with the sodium iodide detector (a) and (b), as well as results of corresponding simulation calculations (c) and (d).

\section{Discussion} \label{discussion}
\noindent The spectrum in green color of Fig. \ref{ExpPeakSearch} (a) exhibits a clear narrow peak with an energy of 1.30 MeV. It appears only at a negative observation angle $\theta_x$ = - 0.2 mrad, while at the positive angle of + 0.2 mrad mainly background is observed, see spectrum in blue colour. The result shows that our method for a peak search, as proposed in \cite[Fig. 11]{BacL25} and applied for an unsuccessful experiment, demonstrates that our method is highly sensitive and enables the creation of a difference spectrum shown in Fig. \ref{ExpPeakSearch} (b). This crucial methodical procedure has not been addressed in \cite{PavK20,PavK21}, however, features of the experimental observations were predicted.
In contrast to the above mentioned Si$_{1-x}$Ge$_x$ strained layer crystal, the diamond undulator structure cannot be separated from the host diamond crystal. As a consequence, one is always faced with radiation from the backing, the thickness of which can be minimized to values in the several $\mu$m range. However, in our experiment the thickness in [100] direction was 182 $\mu$m, a rather large value in comparison with the undulator layer thickness of 14.1 $\mu$m.
Therefore, a huge channeling radiation contribution is present, although somehow suppressed due to the choice of the observation direction, and consequently the difference spectrum is afflicted with large uncertainties, in particular above 2 MeV. The thickness of the substrate can be substantially reduced by nearly two orders of magnitude into the several $\mu$m region. This can be achieved by laser water jet cutting, followed by quantum polishing to smooth the back surface.

Figs. \ref{ExpPeakSearch} (c) and (d)  depict Monte-Carlo simulated spectra under idealized conditions. The calculations are based on the formalism described in appendix \ref {simulation}. The bremsstrahlung contribution was estimated from \cite[Fig. 9 (c)]{BacL25} to be small and has been neglected. With some care in mind, the simulated spectrum (d) will be compared with the experimentally observed one (b). The peak energy from the simulation calculation of 1.28 MeV is close to the measured one of 1.30 MeV. The experimentally observed spectral width is broader, and the total intensity is significantly lower than predicted by simulation. The Bragg-diffraction image of Fig. \ref{SIMSBraggMeasurement} (b) may indicate that the amplitude decreases by relaxation of the stress, probably by misfit dislocations \cite{Bre97}, a fact not taken into account in our simulation calculation. In addition, an infinitesimal small aperture was assumed which is also unrealistic, not to mention that the lattice expansion constant of the modified Vegard's law could be different, and that also the simulation model has systematic errors.

We add that a peak was also observed when the crystal was  aligned into the (111) plane.

For applications one has to cope with the unavoidable and rather intense channeling radiation distribution tail \cite{SusK24}. However, it can be minimized by optimizing the boron doping profile and the number of periods, see Fig. \ref{SimSpectra855MeVlambdaVariable} (b) in appendix \ref{simulation}. To cope with the tail, difference measurements must be performed by de-tuning the photon peak energy or varying the observation angle. While a de-tuning off the resonance under investigation has a drastic effect, a shift of the continuum energy has only a minor effect on the intensity of high energy parasitic resonances.

\section{Outlook} \label{prospects}
As has been shown in the appendix \ref{simulation}, at the MAMI facility with a beam energy of 855 MeV narrow band and highly directional photon beams can be provided in a wide photon energy range. Such beams are useful for detector calibration purposes with tailored undulators. However, for nuclear resonance fluorescence experiments the intensity is too low in contrast to various photonuclear reactions in the giant resonance domain. For the latter photons with energies in the order of 15 MeV are required. According to Eq. (\ref{undulatorEnergy}) an as small as possible undulator period $\lambda_U$ should be chosen to keep the beam energy of the electron accelerator as low as possible. However, from the viewpoint of the peak intensity there exist severe restrictions on the magnitude of $\lambda_U$ and by interference phenomena between the channeling and undulator radiation, as pointed out in appendix \ref{simulation}.
The following estimates are performed for a dedicated 3 GeV accelerator to illuminate figures of merits and limitations of the novel radiation source, free of restrictions imposed by existing facilities like the 1.6 GeV facility of MAMI B or 12 GeV Continuous Electron Beam Accelerator Facility (CEBAF) \cite{AddA24}.

\begin{figure}[tb]
\centering
\hspace*{0 mm}
  \includegraphics[angle=0,scale=0.65,clip]{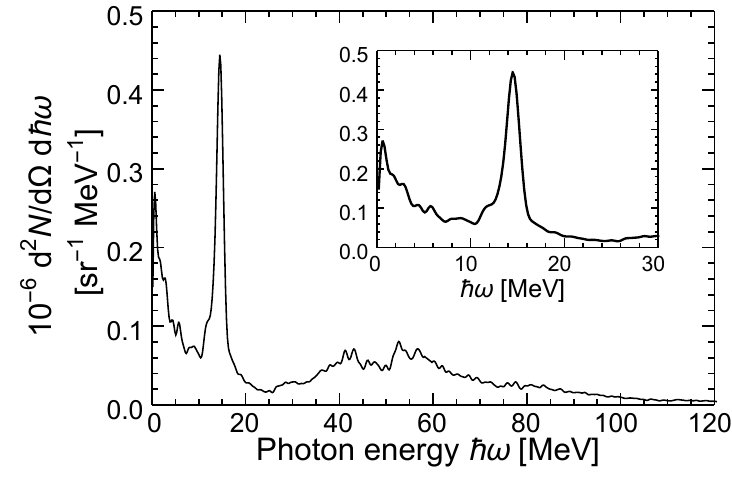}
\caption[]{
Simulated photon spectrum per electron for a sinusoidal boron doping profile at a beam energy of 3.0 GeV. Period length $\lambda_U$ = 5 $\mu$m, minimum barrier height $U_B$ = 9.0 eV,
amplitude $A_U$ = 0.656 \AA, undulator parameter $K$ = 0.48, relaxation length $L_A = \infty$, and number of periods $N_U$ = 12. An infinitesimal small aperture was assumed positioned at $\theta_x = \delta$ = -0.0824 mrad and $\theta_y = 0$. Beam divergence $\sigma_x = \sigma_y$ = 0.06 mrad. Mean of 200 single simulations.} \label{SimPeak3000MeV}
\end{figure}

Results of simulation calculations are shown in Fig. \ref{SimPeak3000MeV}. The main feature is the very intense peak at a photon energy of 14.5 MeV with a full width of half maximum (FWHM) of 1.9 MeV.  The energy integrated peak intensity amounts to about 0.6$\cdot 10^{6}$/sr and electron. Estimations with Eq. (\ref{undulatorEnergy}) show that for $(\gamma\theta)^2\simeq 0.03$ rad$^2 $ the peak energy shift is small compared with its half width. With the fraction of the intensity located in an angular cone with a radius of about 0.03 mrad, the corresponding solid angle is 3~$\cdot 10^{-9}$ sr, the integrated $\gamma$-ray flux amounts to 1.8 $\cdot 10^{-3}$ per electron. At a beam current of 100 $\mu$A totally $1.1 \cdot 10^{12}$ photons per second would be emitted in the peak. The on-target photon flux is orders of magnitudes larger than for Compton-backscattering sources \cite[Table 10]{HowA22}, however, with an high energy photon tail. The principle to cope with it has been described at the end of section \ref{discussion}.

The time structure can be controlled by the electron beam. Some applications may profit from the fact that the emitted $\gamma$-ray beam is expected to be linearly polarized. With such a beam a number of experiments in fields like basic nuclear research, technical applications as well as medicine are conceivable with small sample sizes since even in 10 m distance the $\gamma$- ray beam spot has a radius of only 0.3 mm.

The heat load at a 100 $\mu$m thick diamond target plate at a beam current of 100 $\mu$A can be estimated to be about 10 W which seems to be manageable. The total power of the electron beam amounts to 300 kW.
If energy recovery would be feasible, the power consumption and the radioactive waste in the beam dump can be reduced.


\section{Conclusion} \label{Conclusions}
A four-period diamond superlattice was fabricated via chemical vapor deposition with a sinusoidal boron doping profile.
The accompanied periodic variation of the lattice constant resulted in sinusoidally deformed (110) planes with a period length of 5.0 $\mu$m and an amplitude of 0.098 nm. A channeling experiment was performed with the 855 MeV electron beam of the Mainz Microtron MAMI accelerator facility. A clear peak
was observed with a large sodium iodide scintillation detector at a photon energy of 1.30 MeV. Monte-Carlo simulation calculations, assuming an ideal sinusoidal undulator profile, reproduce well the photon peak energy. However, the measured line width and intensity seem, in comparison, to be broadened and reduced, respectively, for which explanations have been discussed.

With the existing scientific accelerator facility MAMI narrow band photon beams with energies below about 6 MeV could be provided. Higher photon energies require higher electron energies to achieve optimal intensities. Simulation calculations were performed for an intense 14.5 MeV $\gamma$-ray beam with 3 GeV electrons.

The photon energy spectra resemble the radiation characteristics known from the magnetic undulator theory, featuring a strong fundamental peak, but also an unavoidable high-energy tail from channeling radiation. To cope with it, the electron beam energy or the observation direction must be varied.
The high directionality, combined with an intensity orders of magnitude larger than
for Compton-backscattering sources \cite[Table 10]{HowA22} should be beneficial for potential applications with rather small target samples. Many of the fields addressed in \cite{HowA22} are suitable for applications in basic nuclear research, technical applications, or even medicine.

\section*{Data availability statement}
\noindent All data are freely available.
\section*{Acknowledgements} \label{Acknowledgements}
\noindent We appreciated help of Daniel Dominguez, Philipp Kompa, and Yaideny Rodriguez during measurements at MAMI.

\noindent  We thank Andrei V. Korol and Andrey V. Solov'yov for discussions on the undulator design and organizational support of the project.

\noindent This work has been financially supported by the European Innovation Council (EIC) Pathfinder TECHNO-CLS project 101046458 and the N-light project H2020-MSCA-RISE N-LIGHT  No. 872196.

\section*{Authors contributions}
%
%
%
%
%
%

\noindent \textbf{Boron-doped diamond growth:} David Eon

\noindent \textbf{Diamond preparation:} Thu Nhi Tran Caliste

\noindent \textbf{Crystal characterization:} Patrik Straňák, Lutz Kirste, Thu Nhi Tran Caliste, Rébecca Dowek

\noindent \textbf{Model development:} Hartmut Backe, Werner Lauth, Thu Nhi Tran Caliste

\noindent \textbf{ESRF beamtime:} Rébecca Dowek, Pierre Everaere, Simon Bénichou, José Baruchel, Thu Nhi Tran Caliste

\noindent \textbf{MAMI beamtime:} Werner Lauth, Pascal Klag, Rébecca Dowek, Thu Nhi Tran Caliste

\noindent \textbf{Editor:} Hartmut Backe

\section{Appendices} \label{appendices}
\subsection{Superlattice design and fabrication} \label{superlattice}
\noindent The design profile of the boron doping in [100] direction is of sinusoidal shape with minimum and maximum boron content $x_{b,min}$ = $c_{b,min}$/$n_C$ = 5.67 $\cdot 10^{-4}$ and $x_{b,max}$ = $c_{b,max}$/$n_C$  = 56.7 $ \cdot 10^{-4}$. The corresponding boron concentrations are $c_{b,min}$ = 1.0~$\cdot 10^{20}$/cm$^3$ and $c_{b,max}$ = 10.0~$\cdot 10^{20}$/cm$^3$, respectively, the density of carbon atoms for diamond is $n_C = 1.763 \cdot10^{23}$/cm$^3$. As function of the $z$ coordinate the boron content reads for the first half period $x_b(z) = x_{b,min}+(x_{b,max}-x_{b,min})/2 \cdot (1-\cos[2\pi z/\lambda_U])$, $0\leq z\leq \lambda_U/2$. The continuation to $\lambda_U/2< z\leq \lambda_U$ can be obtained by symmetry considerations.
\begin{figure}[b]
\centering
    \includegraphics[angle=0,scale=0.4,clip]{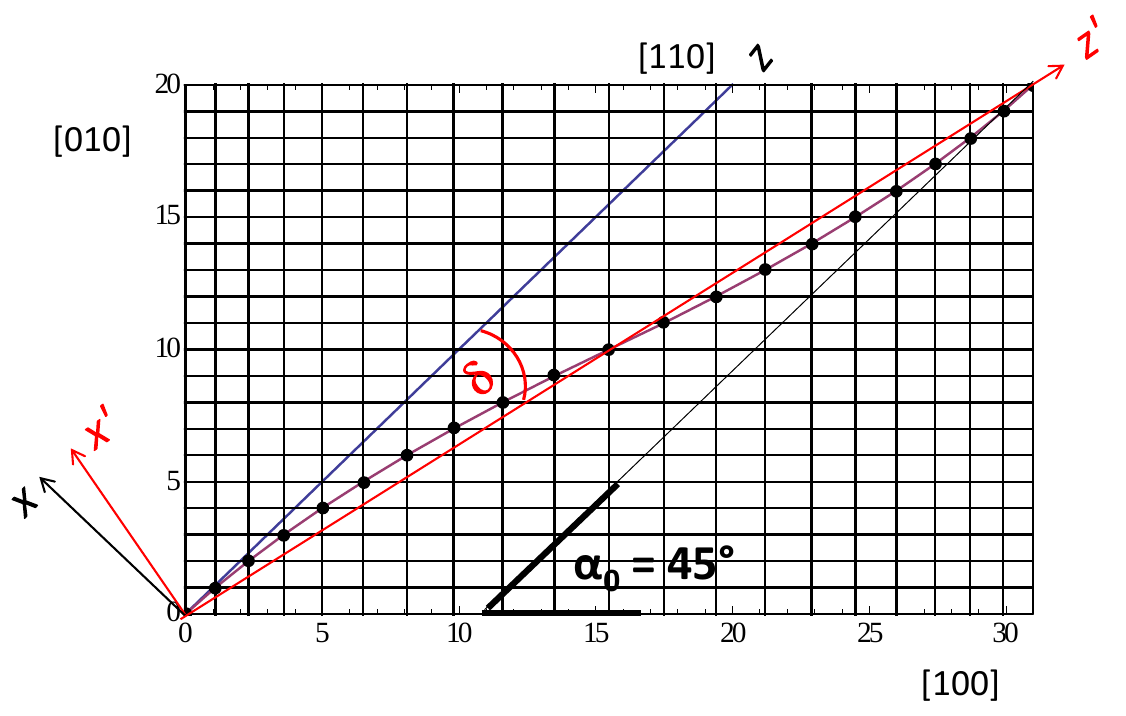}
\caption[]{Effect of a periodic variation of the lattice constant. A superlattice (undulator) crystal originates as shown by the full line with black dots for one period. The effect of the tetragonal distortion of the equilibrium cubic unit cell has been described by M.B.H. Breese \cite{Bre97}. The electron beam enters in $z$ direction with an angle $\delta_0$ = - 0.0272 mrad with respect of the [110] direction of the plane backing crystal. The $z'$ direction is the undulator direction. The two coordinate systems $(x,z)$ and the $(x',z')$ are interconnected by a rotation transform with an angle $\delta$ = - 0.150 mrad. Planes are formed perpendicular to the plane of drawing.  Figure taken from \cite{BacL25}.
}\label{UndulatorCrystal}
\end{figure}

The lattice expansion for the "atomic volume interpolation" \cite[Fig. 1]{WojA08}, also called linear  "modified Vegard's low" reads $\Delta d_{110}/d_{110}(z) = k \cdot x_b$ with the slope constant $k = 0.0962$. A differential equation $dx_b/dz = -k \cdot x_b(z/\sqrt{2})/2$ must be solved to obtain the nominal bending of the (110) channel with $x_b(z\rightarrow0) = 0$. The angle at the entrance is $\delta_0$ = $dx_b/dz(z\rightarrow0)$ = - 0.0272 mrad, caused by the residual boron doping. The solution is schematically shown in Fig. \ref{UndulatorCrystal}.

A periodically bent (110) channel results with $\lambda_U = \sqrt{2}\cdot$ 3.54 $\mu$m = 5.0 $\mu$m which is schematically shown in Fig. \ref{UndulatorCrystal}.
It defines the $(x^\prime,z^\prime)$ coordinate system with the $z^\prime$ axis through the turning points of the derivative, rotated clock wise by an angle $\delta = \delta_0-\delta_U$ = - 0.150 mrad, $A_U$ = 0.977 \AA, $\delta_U = 2\pi/\lambda_U \cdot A_U$ = 0.123 mrad, with respect to the nominal [110] direction of the un-doped diamond backing crystal.

When fabricating undulator devices attention must be paid to both the quality of the doped overlayers, and the quality of diamond substrate. Our application requires single-crystal diamond substrates with the main surface orientation (100), with a miscut angle of 2 degrees within one surface, and with highly perfect bulk and surface crystal quality. In addition, such diamond plates should be available with large areas (several mm$^2$) free of extended defects such as dislocations. The diamond surface plays also an important role for the quality of the overgrown layers on it, therefore it must be as flat as possible. With special mechanical- chemical polishing, we reach the flat surface of less than 1nm r.m.s.
Our final choice is high-crystalline-quality HPHT (High Pressure High Temperature) type IIa substrates with the size around 5x5 mm$^2$. The thickness of the diamond substrate varies from 80 to 270 µm from one side to the other, due to the miscut angle of 2 degrees. The Diffraction Rocking Curve Imaging (RCI) technique shows not many defects present in the bulk: some dislocations, no stacking faults, no growth sector boundaries, no inclusion and very few defects introduced by surface processing.

For the growth of the boron doped diamond layers, the diamond sample was first subjected to a tri-acid cleaning process consisting of 1 volume of HClO$_4$, 4 volumes of HNO$_3$, and 3 volumes of H$_2$SO$_4$ for 1 hour under boiling conditions. This step effectively removes residual graphitic surface layers. The sample was then rinsed with acetone, followed by alcohol and isopropyl alcohol (IPA). Subsequently, the sample was loaded into a microwave-assisted CVD (Chemical Vapor Deposition) reactor for diamond growth. The reactor plasma was ignited under a hydrogen atmosphere at a pressure of 44 mbar and a power of 250 W. The sample was exposed to this plasma for 1 hour to ensure uniform temperature distribution across both the sample and the reactor interfaces.

The growth process itself is a key innovation to achieve the desired doping profile required for the specific application. A tailored dilution stage, with precise control of the mass flow controllers, was employed to ensure a sinusoidal boron doping pattern in the gas phase. During the growth phase, a gas flow of 100 sccm of H$_2$ and 4 sccm of CH$_4$ was maintained, with a variable B$_2$H$_6$ concentration injected into the gas mixture. Based on established calibration curves, the expected boron doping level in the diamond ranges from 1 $ \cdot10^{20}$ cm$^{-3}$  to 10 $ \cdot 10^{20}$ cm$^{-3}$. Temperature control was achieved using a pyrometer, maintaining an average temperature of 850°C throughout the process. Considering the growth rates, each doping cycle lasted for 6015 seconds with the growth rate of 0.58 nm/s.

\subsection{Monte-Carlo simulation calculations} \label{simulation}
\noindent
The Monte-Carlo simulation code described in \cite{BacL25} holds only for triangular doping profiles resulting in a constant bending radius for each half period with opposite signs. For this work, therefore, it was reformulated for sinusoidal shapes avoiding approximations.

The starting point is the equation of motion \cite[Eq. (2.78)]{BirCh97} in the non-inertial undulating reference frame
\begin{eqnarray}
\label{eqBiryukov}
pv\frac{dx'^2}{dz'^2}=-\frac{du(x')}{dx'}-\frac{pv}{R(z')}=-\frac{dU(x',z')}{dx'}\\
\label{Upot}
U(x',z')=u(x')+\frac{pv}{R(z')}x'\\
\label{RzInvers}
1/R(z') = -(1/R_{min}) \sin(2\pi z'/\lambda_U).
\end{eqnarray}
In Eq. (\ref{eqBiryukov}) $u(x')$ is the planar (110) potential in the $(x',z')$ coordinate system, defined in Fig. \ref{UndulatorCrystal}, which periodically repeats with the inter-planar distance $d_p$. The additional term $pv/R(z')$ describes according to Eq. (\ref{RzInvers}) the centrifugal force with $R(z')$ the bending radius with $0 < 1/R_{min} = A_U (2\pi/\lambda_U)^2$ its minimum value, $A_U$ is the undulator amplitude. It is a function of the longitudinal coordinate $z'$.
In addition are $pv$ = $\beta^2\gamma~ m_e c^2$ with $m_e c^2$ the rest mass of the electron, $\gamma$ the Lorentz factor,
and $\beta = v/c =\sqrt{1-1/\gamma^2}$ the reduced speed $v$ of the electron.

With the second equal sign in Eq. (\ref{eqBiryukov}) formally a potential $U(x',z')$, Eq. (\ref{Upot}), has been introduced.
A particle captured in the pocket, whose depth varies with $z'$, is steered by it and performs, in addition to the channeling oscillation, the wanted undulator oscillation.

The simulation calculation has been performed for the first interval $\Delta z$ = 100 \AA~by solving numerically Eq. (\ref{eqBiryukov}) with the randomly determined initial values $x(0)$ and $\vartheta_x(0) = dx'/dz'(0)$. At the iteration process the scattering angle change $\Delta\vartheta_x$ by collisions with atoms and also electrons, notice that \cite{BacL25} is based upon \cite{Bac22A} in which both interactions have been treated, was calculated for the $n^{th}$ interval with constant thickness $\Delta z$.  For the $(n+1)^{th}$ interval Eq. (\ref{eqBiryukov}) was solved with the modified initial values $\vartheta_x((n+1) \Delta z) = \vartheta_x(n \Delta z) + \Delta\vartheta_x(\Delta z)$ and $x((n+1) \Delta z) = x(n \Delta z)$, whereby $n$ denotes the exit values of the preceding numerical solution.

Transformation into the inertial $(x,z)$ system is done by a rotation with the angle $0<\delta = A_U 2 \pi/\lambda_U \ll1$ resulting approximately in $x = x' - z' \delta + A_U sin(2 \pi/\lambda_U \cdot z')$ and $z = z'$. The instantaneous angle of the trajectory  must be corrected by a term $ \delta \cdot(1 - cos(2 \pi/\lambda_U \cdot z'))$. Alternatively, the differential equation (\ref{eqBiryukov}) can be transformed into the $(x,z)$ system resulting in
\begin{eqnarray}
\label{eqBiryukovInertial}
pv\frac{dx^2}{dz^2}=-\frac{du(x+z\cdot \delta-A_U sin(2\pi/\lambda_U \cdot z))}{dx}.
\end{eqnarray}
An frequency modulation like term appears which might result in Fig. \ref{SimPeak3000MeV} in the third harmonic and the feature at 56 MeV.

\begin{figure}[b]
\centering
    \includegraphics[angle=0,scale=0.7,clip]{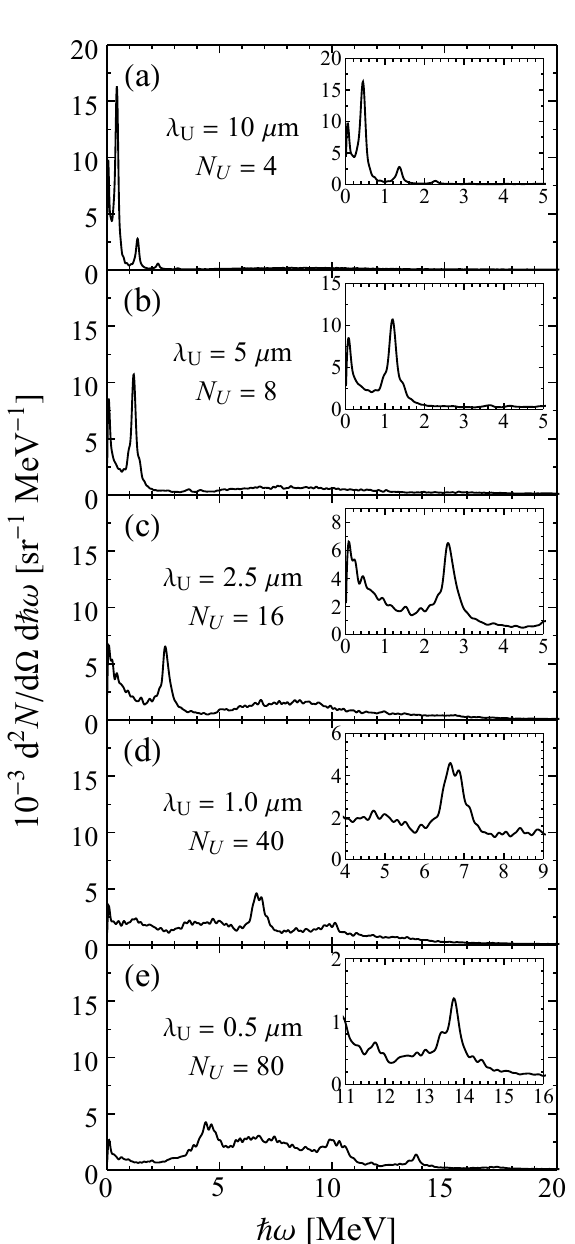}
\vspace*{-0.2cm}
\caption[]{
Simulated photon spectrum per electron for a sinusoidal boron doping profile at a beam energy of 855 MeV. Both, period length $\lambda_U$  and number of periods $N_U$ vary with $\lambda_U N_U$ = 40 $\mu$m a constant, minimum barrier height $U_B$ = 8.5 eV, and minimum bending radius $R_{min}$ = 2.632 mm. Undulator amplitudes $A_U$ are 9.626 \AA ~(a), 2.406 \AA ~(b), 0.602 \AA ~(c), 0.0963 \AA ~(d), 0.0240 \AA ~(e). Undulator parameters $K$ are 1.0130 (a), 0.5063 (b), 0.2531 (c), 0.1013 (d),  0.0506 (e).
Beam divergence $\sigma_x = \sigma_y$ = 0.06 mrad. Mean of 200 single simulations.
}  \label{SimSpectra855MeVlambdaVariable}
\end{figure}

For applications the photon energy $\hbar\omega$ is the primary quantity for which a peak intensity as high as possible, a line width as narrow as possible, and total background radiation as low as possible must be found. In the following the procedure will be sketched to find proper parameters. With Eq. (\ref{Upot}) and (\ref{RzInvers}) a minimum barrier height $U_B = U(x'_{max},\lambda_U/4)-U(x'_{min},\lambda_U/4)$ has been introduced with $x'_{max}$ and $x'_{min}$ both functions of $pv/R_{min}$.
Solving for the latter at a given $U_B$ results in $pv/R_{min} = 2\pi \beta^2 m_e c^2 K/\lambda_U$. The right hand side has been obtained by the replacement $1/R_{min} = A_U (2\pi/\lambda_U)^2$ and elimination of $A_U$ with the undulator parameter $K = \gamma\cdot A_U \cdot 2\pi/\lambda_U$. Since the left hand side is a constant,
it follows that $K$ and $\lambda_U$ are strictly correlated, independent of $\gamma$. Intensities of the peak and the high energy tail are rather sensitive functions of $U_B$. For a somewhat optimized $U_B$ = 8 eV a $pv/R_{min}$ = 0.34 MeV/$\mu$m is obtained, and for a reasonable large $K$ = 0.53 a $\lambda_U$ = 5.0 $\mu$m results. Now $\gamma$ can be calculated with the aid of Eq. (1) for the wanted $\hbar\omega$. The only remaining parameter is the number of periods $N_U$ which must be optimized by simulation calculations. Practical limits originate, in addition to scattering, also from the amplitude relaxation $L_A$ which is a material constant.
Since it is unknown we assumed $L_A = \infty$.

Fig. \ref{SimSpectra855MeVlambdaVariable} shows spectra for a beam energy of 855 MeV at $\lambda_U$ and $N_U$ both variable with $\lambda_U N_U$ = 40 $\mu$m a constant. The peak energy varies with the undulator period as expected, however, at panels (d) and (e) the total background intensity may be for any application too large. The structures may be interpreted as interference phenomena between undulator and channeling radiation. A good compromise may be $\lambda_U$ = 5 $\mu$m and $N_U$ = 8, panel (b), which fixes the photon peak energy somewhere around $\hbar\omega$ = 1.3 MeV. What has been exemplified here holds figuratively for other scenarios, in particular also for the experiments with Si$_{1-x}$Ge$_x$ undulator crystals at a period length of 0.43 - 0.44 $\mu$m  \cite{WisA14,WisU17}.

The code runs on Wolfram Mathematica 14.3, executed on a personal computer. It delivers useful results in reasonable real time of about 10 minutes.


~\newpage
\bibliography{bibfileBa}

\end{document}